\documentstyle[12pt]{article}
\begin{document}
\def\sixj#1#2#3#4#5#6{\left\{\negthinspace\begin{array}{ccc}
#1&#2&#3\\#4&#5&#6\end{array}\right\}}
\def\threej#1#2#3#4#5#6{\left(\negthinspace\begin{array}{ccc}
#1&#2&#3\\#4&#5&#6\end{array}\right)}
\def\mbn{\mbox{\boldmath$\nabla$}}
\def\rbh {\hat{\bf r}}
\def\rb {{\bf r}}
\def\kb {{\bf k }}
\def\qb {{\bf q}}
\def\kr {{\bf k}\cdot{\bf r}}
\def\pb {{\bf p}}
\def\etal{{\it et al., }}
\def\cf{{\it cf }}
\def\ie{{\em i.e., }}
\def\kev{\; {\rm keV} }
\def\mev{\; {\rm MeV} }
\def\ev{\; {\rm eV} }
\def\tev{\; {\rm TeV} }
\def\eV{\; {\rm eV} }
\def\gev{\; {\rm GeV} }
\def\etc{ {\it etc}}
\def\epr {e^{i{\bf p}\cdot{\bf r}}}
\def\qrv {{\bf q}\cdot\hat{{\bf r}}}
\def\prv {{\bf p}\cdot\hat{{\bf r}}}
\def\pdm{\frac{\partial \; }{\partial m^2} }
\def\pdM{\frac{\partial \; }{\partial M^2} }
\def\pdphi{\frac{\partial \; }{\partial \phi} }
\def\eqr {e^{i{\bf q}\cdot{\bf r}}}
\def\dket#1{||#1 \rangle}
\def\dbra#1{\langle #1||}
\def\isim{\:\raisebox{-0.5ex}{$\stackrel{\textstyle.}{=}$}\:}
\def\lsim{\:\raisebox{-0.5ex}{$\stackrel{\textstyle<}{\sim}$}\:}
\def\gsim{\:\raisebox{-0.5ex}{$\stackrel{\textstyle>}{\sim}$}\:}
\def\a {{\alpha}}
\def\b {{\beta}}
\def\e {{\epsilon}}
\def\g {{\gamma}}
\def\r {{\rho}}
\def\s {{\sigma}}
\def\k {{\kappa}}
\def\l {{\lambda}}
\def\m {{\mu}}
\def\n {{\nu}}
\def\r {{\rho}}
\def\t {{\tau}}
\def\w {{\omega}}
\def\be{\begin{equation}}
\def\ee{\end{equation}}
\def\br{\begin{eqnarray}}
\def\er{\end{eqnarray}}
\def\brn{\begin{eqnarray*}}
\def\ern{\end{eqnarray*}}
\def\x{\times}
\def\go{\rightarrow  }
\def\rf#1{{(\ref{#1})}}
\def\nn{\nonumber }
\def\pd#1#2{\frac{\partial #1}{\partial #2} }
\def\ket#1{|#1 \rangle}
\def\bra#1{\langle #1|}
\def\Ket#1{||#1 \rangle}
\def\Bra#1{\langle #1||}
\def\ov#1#2{\langle #1 | #2  \rangle }
\def\hf {{1\over 2}}
\def\hw {\hbar \omega}
\def\mbs{\mbox{\boldmath$\sigma$}}
\def\gsim{\:\raisebox{-0.5ex}{$\stackrel{\textstyle>}{\sim}$}\:}
\def\lsim{\:\raisebox{-0.5ex}{$\stackrel{\textstyle<}{\sim}$}\:}
\def\pdM{\frac{\partial \; }{\partial M^2_{\pm} }}
\def\pdrnm{\frac{\partial \; }{\partial r_{nm}} }
\def\Ket#1{||#1 \rangle}
\def\Bra#1{\langle #1||}
\def\mbn{\mbox{\boldmath$\nabla$}}
\def\sss{\scriptscriptstyle}
\def\ss{\scriptstyle}
\def\gp{g_{\sss +}}
\def\gm{g_{\sss -}}
\def\endauthors{}
\def\authors#1\endauthors{#1}
\def\pdMpm{\frac{\partial \; }{\partial M^2_{\pm} }}
\def\ninj#1#2#3#4#5#6#7#8#9{\left\{\negthinspace\begin{array}{ccc}
#1&#2&#3\\#4&#5&#6\\#7&#8&#9\end{array}\right\}}
\begin{titlepage}
\pagestyle{empty}
\baselineskip=21pt
\rightline{McGill/96-36}
\rightline{hep-ph/9609421}
\vskip .2in
\begin{center}
{\large{\bf Exact evaluation of the nuclear form factor for new kinds of
majoron emission in neutrinoless double beta decay}}
$^*$
\end{center}

\vskip .1in

\authors
\centerline{C.~Barbero${}^{ a\dagger}$, J.~M.~Cline${}^b$,
F.~Krmpoti\'{c}$^{a\dagger}$ and D.~Tadi\'c${}^c$}
\vskip .15in
\centerline{\it ${}^a$ Departamento de F\'\i sica, Facultad de Ciencias
Exactas}
\centerline{\it
Universidad Nacional de La Plata, C. C. 67, 1900 La Plata, Argentina.}
\vskip .1in
\centerline{\it ${}^b$ McGill University, Montr\'eal, Qu\'ebec H3A 2T8,
Canada.}
\vskip .15in
\centerline{\it ${}^c$ Physics Department, University of Zagreb}
\centerline{\it Bijeni\v cka c. 32-P.O.B. 162, 41000 Zagreb, Croatia.}
\endauthors
\vskip 0.5in
\centerline{ {\bf Abstract} }
\baselineskip=16pt
We have developed a formalism, based on the Fourier-Bessel expansion,
that facilitates the evaluation of matrix elements involving nucleon
recoil operators, such as appear in serveral exotic forms of
neutrinoless double beta decay ($\beta\beta_{0\nu}$).  The method is
illustrated by applying it to the ``charged'' majoron model, which is
one of the few that can hope to produce an observable effect.  From our
numerical computations within the QRPA performed for $^{76}Ge$,
$^{82}Se$, $^{100} Mo$, $^{128}Te$ and $^{150}Nd$ nuclei, we test the
validity of approximations made in earlier work to simplify the new
matrix elements, showing that they are accurate to within 15\%.  Our
new method is also suitable for computing other previously unevaluated
$\beta\beta_{0\nu}$ nuclear matrix elements.
\bigskip

\vspace{0.5in}
\noindent
$^*$Work partially supported by the Fundaci\'on Antorchas, Argentina,
the CONICET from Argentina, the Universidad Nacional de La Plata, NSERC
of Canada and FCAR of Qu\'ebec.\\ $^{\dagger}$Fellow of the CONICET
from Argentina.
\end{titlepage}
\baselineskip=18pt

Since the classic majoron model of Gelmini and Roncadelli \cite{Gel81}
was excluded by the LEP measurement of the invisible width of the $Z$
boson \cite{Abr89,Gon89}, several new possibilities have been proposed
for generating exotic neutrinoless double-beta decay with the
emission of a massless majoron ($\beta\beta_{M}$).  An interesting
class of such models are those where the majoron is either a scalar
\cite{Bur93} or a vector \cite{Car93} boson that has lepton number $-2$
to balance that of the emitted electrons.  These ``charged majoron"
models (CMM's), so-called because the majoron carries the unbroken
$U(1)$ charge of lepton number, are probably the only ones which have a
chance of producing $\beta\beta_{M}$ at a rate which could be observed
in the present generation of experimental searches
\cite{Bam95,Bar96,Hir96}.

>From the point of view of nuclear physics, the CMM's are interesting
because their predicted rate of $\beta\beta_{M}$ depends on different
matrix elements than in previously considered types of double beta
decay, which had not been hitherto computed.  In a previous publication
\cite{Bar96} we made a first step toward computing the CMM matrix
elements by neglecting a number of operators, an approximation which,
while widely used in the literature, should be validated by more
accurate computations.  The aim of this letter is twofold. First, we
elaborate a formalism, based on the Fourier-Bessel expansion introduced
in ref.~\cite{Krm92}, that
allows us to evaluate in a rather simple way the matrix elements of
nuclear recoil operators, such as appear in the CMM.  Second, to
quantify the contributions of the operators neglected previously, we
perform the corresponding numerical calculations in the QRPA for
$^{76}Ge$, $^{82}Se$, $^{100} Mo$, $^{128}Te$ and $^{150}Nd$ nuclei.

We begin by describing the matrix elements of interest and the
approximations that were previously made.  In the CMM of
ref.~\cite{Bur93}, on which we will focus in this letter, the inverse
half-life for $\beta\beta_{M}$-decay is
\begin{equation}
[T(0^+_i\rightarrow 0^+_f)]^{-1}=g_{\sss CM}^2 {\cal G}_{\sss CM}
|{\cal M}^+_{\sss CM} - {\cal M}^-_{\sss CM}|^2,
\label{1} \end{equation}
where $g_{\sss CM}$ is the effective majoron coupling defined in
ref.~\cite{Bar96},
\begin{equation}
{\cal G}_{\sss CM}= \frac{(G_F\cos\theta_C)^4}{128\pi^7\ln 2}
\int(Q-\epsilon_1-\epsilon_2)^3\prod_{i=1}^2 k_i\epsilon_i F(\epsilon_i)
d\epsilon_i,
\label{2} \end{equation}
is the kinematical factor in natural units \cite{Doi88} and
\begin{equation}
{\cal M}_{\sss CM}^{\pm}=\frac{1}{4\pi} \bra{0_f^+}\sum_{nm}
\int k^2dk v(k;M_{\pm}) {\sf O}_{CM}(k,\rb_{nm}) \ket{0_i^+},
\label{3} \end{equation}
with
\be
{\sf O}_{CM}(k,\rb_{nm})=
i\int d\Omega_k \kb\cdot{\bf Y}_{Rnm}e^{i\kr_{nm}},
\label{4}\ee
are the nuclear form factors for charged majoron emission corresponding to
the exchange of two neutrinos with masses $M_{\sss \pm}$ \cite{Bar96}.
Here
\be
v(k;M_{{\sss} \pm})
=\frac{2}{\pi}\left\{\frac{1}{M_{{\sss} \pm}^2}\left[ \frac{1}
{\omega_{{\sss} \pm}(\omega_{{\sss} \pm}+\mu)}- \frac{1}{k(k+\mu)}\right]+
\frac12\pdMpm\frac{1}{\omega_{{\sss} \pm} (\omega_{{\sss} \pm}+\mu)} 
\right\},
\label{5}\ee
$\omega_{\sss \pm}=\left(k^2+M_{\sss \pm}^2\right)^{\hf}$, and $\mu$ is the
average excitation energy of the intermediate nuclear states. The operator
\begin{equation}
\label{6}
{\bf Y}_{Rnm}=i\left[ g_{\sss A}^2 (\mbs_nC_m-C_n\mbs_m)
+ig_{\sss V}g_{\sss A} (\mbs_n\times{\bf D}_m+ {\bf D}_n\times\mbs_m)
+g_{\sss V}^2({\bf D}_n-{\bf D}_m)\right],
\end{equation}
with
\br
C_n&=&\frac{1}{2M}\mbs_n\cdot(2\pb_n+\qb_n),\nonumber\\
{\bf D}_n &=&\frac{1}{2M}(2\pb_n+\qb_n +if_{\sss W} \mbs_n\times\qb_n),
\label{7}\er
is the impulse non-relativistic approximation for the recoil term,
obtained through the standard Foldy-Wouthuysen transformation
\cite{Ros54}-\cite{Com83}.   $M$ is the
nucleon mass, $f_{\sss W}=4.7$ is the effective weak-magnetism coupling
constant and $\pb_n\equiv-i\mbn^N_n$ and $\qb_n\equiv-i\mbn^L_n $ are,
respectively, the nuclear and lepton operators acting at the point
$\rb_n$, with the understanding that $\mbn^N_n$ operates only on the
nucleon wave functions, while $\mbn^L_n$ acts only on the exponential
factor in eq.~(\ref{4}).  Therefore one can make the replacements
\be
\qb_n\go \kb,~~~~ \qb_m\go -\kb,
\label{8}\ee
which yields\footnote{In eqs. \rf{7} and \rf{9} we have neglected the 
induced
pseudoscalar interaction term. It is relatively small and can be easily
embodied in eq. \rf{9} by simply multiplying the operator
$g_{\sss A}^2(\mbs_n\cdot\kb)(\mbs_m\cdot\kb)$ by the factor
$(1+h_{\sss P}W_0/2)$, where
$h_{\sss P}=g_{\sss P}/g_{\sss A}\cong 2M/m_{\pi}^2\sim 1/20$, and
$W_0$ is $\beta\beta$ transition energy of the order of $\mev$.}
\br
iM\kb\cdot{\bf Y}_{Rnm}&=&-g_{\sss A}^2\left[
-(\mbs_n\cdot\kb)(\mbs_m\cdot\kb)
+(\mbs_n\cdot\kb)(\mbs_m\cdot\pb_m) -(\mbs_n\cdot\pb_n)(\mbs_m\cdot\kb)
\right]
\nonumber\\
&-&g_{\sss V}g_{\sss A}\left[
i\kb\cdot (\mbs_n\times\pb_m +\pb_n\times\mbs_m)
+f_{\sss W}[(\mbs_n\cdot\kb)(\mbs_m\cdot\kb)-k^2(\mbs_n\cdot\mbs_m)\right]
\nonumber\\
&-&g_{\sss V}^2\left[ (\pb_n-\pb_m)\cdot\kb+k^2 \right].
\label{9}
\er

In our previous work \cite{Bar96} we adopted the approximation
made in another context by Doi \etal \cite{Doi88}\footnote{Our relative
signs differ in this equation and in eq.~(\ref{6}) from the corresponding 
ones
of ref.~\cite{Doi88} because of a sign error in the latter.}
\be
i\kb\cdot{\bf Y}_{Rnm}\cong g_{\sss A}\frac{k^2}{3M}\mbs_n\cdot\mbs_m
( 2g_{\sss V}f_{\sss W}+g_{\sss A}),
\label{10}
\ee
and which was also used by ref.~\cite{Hir96}.  Although a series of
different justifications was given for this procedure, and we will
expand upon them below, operationally it means to neglect in \rf{9} all
terms containing the nucleon momenta, as well as the term $g_{\sss
V}^2k^2$, and to use the relation
\be
(\mbs_n\cdot\kb)(\mbs_m\cdot\kb)\cong \frac{1}{3}k^2
\mbs_n\cdot\mbs_m.
\label{11}\ee
However the nuclei momenta are necessarily of the same order as the
momentum transfer $k$, and from this point of view the trustworthiness
of the estimate \rf{10} is questionable and it should be checked
numerically.  In the remainder of this letter we shall describe a
method for so doing, and illustrate its use by computing for several
nuclei the contributions to the matrix elements of $\kb\cdot{\bf
Y}_{Rnm}$ that were previously neglected.

To begin, we rewrite the operator ${\sf O}_{CM}(k,\rb_{nm})$ in the form
\be
{\sf O}_{CM}(k,\rb_{nm})= {\sf O}_{CM}^{\sss AA}(k,\rb_{nm})+
{\sf O}_{CM}^{\sss AV}(k,\rb_{nm})+
{\sf O}_{CM}^{\sss VV}(k,\rb_{nm}),
\label{12}\ee
with
\br
{\sf O}_{\sss CM}^{\sss AA}(k,\rb_{nm})&=&\frac{g_{\sss A}^2}{M}
\int d\Omega_k e^{i\kr_{nm}}
[(\mbs_n\cdot\kb)(\mbs_m\cdot\kb) -2(\mbs_n\cdot\kb)(\mbs_m\cdot\pb_m)],
\nonumber\\
{\sf O}_{\sss CM}^{\sss AV}(k,\rb_{nm})&=& -\frac{g_{\sss A}g_{\sss V}}{M}
\int d\Omega_k e^{i\kr_{nm}}
\left\{i2\kb\cdot \mbs_n\times\pb_m
+f_{\sss W}[(\mbs_n\cdot\kb)(\mbs_m\cdot\kb)-k^2(\mbs_n\cdot\mbs_m)]\right\}
\nonumber\\
{\sf O}_{\sss CM}^{\sss VV}(k,\rb_{nm})&=&-\frac{g_{\sss V}^2}{M}
\int d\Omega_k e^{i\kr_{nm}} (2\pb_n\cdot\kb+k^2).
\label{13}\er
In deriving the last result we used the fact that the terms containing
$\pb_m$ contribute in the same way as the analogous terms containing
$\pb_n$, \ie  the contribution of $\kb\cdot (\mbs_n\times\pb_m)$ is
equal to that of $-\kb\cdot (\mbs_m\times\pb_n)$, {\it
etc}.\footnote{This can be easily shown from the multipole expansion
that follows.} To obtain the corresponding matrix elements
\begin{equation}
{\cal M}_{\sss CM}^{\sss X\pm}=\frac{1}{4\pi} \bra{0_f^+}
\sum_{nm}
\int k^2dk \,v(k;M_{\pm}) {\sf O}_{CM}^{\sss X}(k,\rb_{nm}) \ket{0_i^+},
\label{14} \end{equation}
where
$\ket{0_i^+}$ and
$\ket{0_f^+}$ are the initial and final nuclear states, respectively, and
${\ss X}$ stands for ${\ss AA}$, ${\ss AV}$ and ${\ss VV}$,
we do the following.
First, we
perform the multipole expansion in \rf{13} and integrate over the angular
coordinates $d\Omega_k$. Second, we
use the angular momenta algebra and introduce a complete set of
intermediate states $\ket{J^{\sss\pi}_{\sss\alpha}}$. In this way we get
\br
&&{\cal M}_{\sss CM}^{\sss AA\pm}=\frac{4\pi g_{\sss A}^2}{M}\int dk k^3 
v(k;M_{\pm})\sum_{LL'J^{\sss\pi}_{\sss\alpha}}
i^{L+L'}\nonumber\\
&\x&\left\{
2 \delta_{LL'}i^{L+J}(-1)^J\hat{L}\hat{J}^{-1}(L1|J)
\dbra{0^{\sss +}_{\sss f}}{\sf T}_{LJ}(k,\rb,\mbs)
\dket{J^{\sss\pi}_{\sss\alpha}}\dbra{J^{\sss\pi}_{\sss\alpha}}
{\sf Y}_{J}(k,\rb,\mbs\cdot\pb)\dket{0^{\sss +}_{\sss i}}\right.
\nonumber\\
&-&\left.k\hat{L}\hat{L'}\hat{J}^{-2}
(L1|J)(L'1|J)
\dbra{0^{\sss +}_{\sss f}}{\sf T}_{LJ}(k,\rb,\mbs)
\dket{J^{\sss\pi}_{\sss\alpha}}
\dbra{J^{\sss\pi}_{\sss\alpha}}{\sf T}_{L'J}(k,\rb,\mbs)
\dket{0^{\sss +}_{\sss i}}
\right\},
\nonumber\\
\label{15}\er
\br
&&{\cal M}_{\sss CM}^{\sss AV\pm}=
\frac{4\pi g_{\sss A}g_{\sss V}}{M}\int dk k^3 
v(k;M_{\pm})\sum_{LL'J^{\sss\pi}_{\sss\alpha}}i^{L+L'}
\nonumber\\
&\x&\left\{
-f_{\sss W} k\delta_{LL'}
\dbra{0^{\sss +}_{\sss f}}{\sf T}_{LJ}(k,\rb,\mbs)
\dket{J^{\sss\pi}_{\sss\alpha}}\dbra{J^{\sss\pi}_{\sss\alpha}}
{\sf T}_{LJ}(k,\rb,\mbs)\dket{0^{\sss +}_{\sss i}}\right.
\nonumber\\
&+&f_{\sss W}k\hat{L}\hat{L'}\hat{J}^{-2}
(L1|J)(L'1|J)
\dbra{0^{\sss +}_{\sss f}}{\sf T}_{LJ}(k,\rb,\mbs)
\dket{J^{\sss\pi}_{\sss\alpha}}\dbra{J^{\sss\pi}_{\sss\alpha}}
{\sf T}_{L'J}(k,\rb,\mbs)\dket{0^{\sss +}_{\sss i}}\nonumber\\
&-&\left.2(-1)^{L+J}\sqrt{6}\hat{L} \sixj{L}{J}{1}{1}{1}{L'}
(L1|L')
\dbra{0^{\sss +}_{\sss f}}{\sf T}_{LJ}(k,\rb,\mbs)
\dket{J^{\sss\pi}_{\sss\alpha}}
\dbra{J^{\sss\pi}_{\sss\alpha}}{\sf T}_{L'J}(k,\rb,\pb)
\dket{0^{\sss +}_{\sss i}}\right\},
\nonumber\\
\label{16}\er
\br
&&{\cal M}_{\sss CM}^{\sss VV\pm}=\frac{4\pi g_{\sss V}^2}{M}\int dk k^3 
v(k;M_{\pm})\sum_{LJ^{\sss\pi}_{\sss\alpha}}
\nonumber\\
&\x&\left\{-k\delta_{LJ}\right.(-1)^J
\dbra{0^{\sss +}_{\sss f}}{\sf Y}_{J}(k,\rb,1)
\dket{J^{\sss\pi}_{\sss\alpha}}\dbra{J^{\sss\pi}_{\sss\alpha}}
{\sf Y}_{J}(k,\rb,1)\dket{0^{\sss +}_{\sss i}}\nonumber\\
&-&2 i^{L+J}\hat{L}\hat{J}^{-1}(L1|J)
\left.\dbra{0^{\sss +}_{\sss f}}{\sf Y}_{J}(k,\rb,1)
\dket{J^{\sss\pi}_{\sss\alpha}}\dbra{J^{\sss\pi}_{\sss\alpha}}
{\sf T}_{LJ}(k,\rb,\pb)\dket{0^{\sss +}_{\sss i}}\frac{}{}\right\}.
\label{17}\er
where $\hat{J}\equiv \sqrt{2J+1}$, $ (L1|J)$ is a short notation for the
Clebsh-Gordon coefficient  $(L010|J0)$ and
\br
{\sf Y}_{LM}(k,\rb,1)&=&j_L(kr)Y_{LM}({\hat{\rb}}),
\nonumber\\
{\sf Y}_{LM}(k,\rb,\mbs\cdot\pb)&=&j_L(kr)Y_{LM}({\hat{\rb}})
(\mbs\cdot\pb),
\nonumber\\
{\sf T}_{LJM}(k,\rb,\mbs)&=&j_L(kr)
(\mbs\otimes Y_{L}({\hat{\rb}}))_{JM},
\nonumber\\
{\sf T}_{LJM}(k,\rb,\pb)&=&j_L(kr)
(\pb\otimes Y_{L}({\hat{\rb}}))_{JM},
\label{18}\er
are spherical one-body operators \cite{Ros54}, the reduced matrix elements
of which appear in eqs.~(\ref{15}-\ref{17}). The explicit dependence
on the energies $\omega _{J^{\sss\pi}_{\sss\alpha}}$
of the intermediate states $\ket{J^{\sss\pi}_{\sss\alpha}}$ in the latter
equations 
can be restored, if desired, by  replacing $\mu\go
\omega _{J^{\sss\pi}_{\sss\alpha}}$ in eq. \rf{5}.
The total matrix element is clearly
\begin{equation}
{\cal M}_{\sss CM}^{\pm}=
{\cal M}_{\sss CM}^{\sss AA\pm}
+{\cal M}_{\sss CM}^{\sss AV\pm}
+{\cal M}_{\sss CM}^{\sss VV\pm},
\label{19} \end{equation}
and in the approximation of eq.~\rf{10} this simplifies to
\br
{\cal M}_{\sss CM}^{\pm}&\cong&
\frac{4\pi g_{\sss A}(2g_{\sss V}f_{\sss W}+g_{\sss A})}{3M}\int dk k^4
v(k;M_{\pm})\sum_{LJ^{\sss\pi}_{\sss\alpha}}(-1)^{L+1}
\nonumber\\
&\x&\dbra{0^{\sss +}_{\sss f}}{\sf T}_{LJ}(k,\rb,\mbs)
\dket{J^{\sss\pi}_{\sss\alpha}}\dbra{J^{\sss\pi}_{\sss\alpha}}
{\sf T}_{LJ}(k,\rb,\mbs)\dket{0^{\sss +}_{\sss i}}.
\label{20}\er
In what follows we shall compute both the the full matrix element and
this approximation in order to compare the two.

The results derived so far are valid for any nuclear model; it remains
only to evaluate the reduced matrix elements $\Bra{0_f^+}{\cal
O}_{LJ}(k,\rb) \Ket{J^{\sss\pi}_{\sss\alpha}}$ and
$\Bra{J^{\sss\pi}_{\sss\alpha}}{\cal O}'_{LJ}(k,\rb)\Ket{0_i^+}$ of the
operators ${\cal O}_{LJ}(k,\rb)$ and ${\cal O}'_{LJ}(k,\rb)$ listed in
\rf{18}.  Within the QRPA formulation, after solving the BCS equations
for the intermediate nucleus \cite{Krm96}, the transition matrix
elements become
\br
\dbra{0_f^+}{\cal O}_{LJ}(k,\rb)\dket{J^{\sss\pi}_{\sss\alpha}}
&=&-\sum_{pn} \dbra{p}{\cal O}_{LJ}(k,\rb)\dket{n}
\left[ v_pu_nX_{pn;\a J}+u_pv_nY_{pn;\a J}\right],
\nonumber\\
\dbra{J^{\sss\pi}_{\sss\alpha}}{\cal O}'_{LJ}(k,\rb)\dket{0_i^+}&=&
-\sum_{pn} \dbra{p}{\cal O}'_{LJ}(k,\rb)\dket{n}
\left[u_pv_nX_{pn;\a J}+ v_pu_nY_{pn;\a J}\right],
\label{21}\er
and the BCS approximation results from the substitution:
\br
&&\sum_{\a}\dbra{0_f^+}{\cal O}_{LJ}(k,\rb)\dket{J^{\sss\pi}_{\sss\alpha}}
\dbra{J^{\sss\pi}_{\sss\alpha}}{\cal O}'_{LJ}(k,\rb)\dket{0_i^+}
\nonumber\\
&&\go\sum_{pn} \dbra{p}{\cal O}_{LJ}(k,\rb)\dket{n}
\dbra{p}{\cal O}'_{LJ}(k,\rb)\dket{n} u_pv_n v_pu_n.
\label{22}\er

For harmonic oscillator radial wave functions, the reduced
single-particle $pn$ form factors are \cite{Ros54,Def66}
\br
\dbra{p}Y_{J}(k,\rb,1)\dket{n}&=&(4\pi)^{-\hf}W(pn;J0J)R^0(pn;J,k),
\nonumber\\
\dbra{p}T_{LJ}(k,\rb,\mbs)\dket{n}&=&(4\pi)^{-\hf}W(pn;L,1,J)R^0(pn;L,k),
\nonumber\\
\dbra{p}T_{LJ}(k,\rb,\pb)\dket{n}&=&(4\pi)^{-\hf}
\left[W_1^{(-)}(pn;L,J) R^{(-)}(pn;L,k) \right.
\nonumber\\
&+&\left.W_1^{(+)}(pn;L,J)R^{(+)}(pn;L,k)\right],
\nonumber\\
\dbra{p}Y_{J}(k,\rb,\mbs\cdot\pb)\dket{n}&=&(4\pi)^{-\hf}
\left[W_2^{(-)}(pn;J) R^{(-)}(pn;J,k)\right.
\nonumber\\
&+&\left.W_2^{(+)}(pn;J)R^{(+)}(pn;J,k)\right],
\label{23}\er
with the angular parts\footnote{We use here the angular momentum coupling 
$\ket{(\hf,l)j}$.}
\br
W(pn;L,S,J)&=&\sqrt{2}\hat{S} \hat{J}\hat{L}\hat{l}_n\hat{j}_n\hat{j}_p
(l_nL|l_p) \ninj{l_p}{\hf}{j_p}{L}{S}{J}{l_n}{\hf}{j_n},
\nonumber\\
W_1^{(\pm)}(pn;L,J)&=&\mp i(-1)^{l_p+j_n+J+\hf} \hat{J}\hat{L}\hat{l}_p
\hat{j}_p\hat{j}_n (l_n+{\ss\hf}\mp {\ss\hf})^{\hf}(l_pL|l_n\mp 1)
\nonumber\\
&\x&\sixj{l_p}{j_p}{\hf}{j_n}{l_n}{J} \sixj{L}{J}{1}{l_n} {l_n\mp 1}{l_p},
\nonumber\\
W_2^{(\pm)}(pn;J)&=&\mp i (-1)^{l_n+j_n+J+\hf}\sqrt{6} \hat{J}\hat{l}_p
\hat{j}_p\hat{j}_n (l_n+{\ss \hf}\mp {\ss\hf})^{\hf}(l_pJ|l_n\mp 1)
\nonumber\\
&\x&\sixj{1}{\hf}{\hf}{j_n}{l_n}{l_n\mp 1} \sixj{l_n\mp 1}{j_n}{\hf}
{j_p}{l_p}{J},
\label{24}\er
and the radial parts
\br
R^0(pn;L,k)&\equiv&
R^L(k;l_p,n_p,l_n,n_n)=
\int_0^\infty u_{n_p,l_p}(r)u_{n_n,l_n}(r)j_L(kr)r^2 dr
\nonumber\\
R^{(\pm)}(pn;L,k)
&=&\pm\left(\frac{\nu}{2}\right)^{\hf}
\left\{(2l_n+2n_n+2\mp 1)^{\hf}R^L(k;l_p,n_p,l_n\mp 1,n_n)\right.
\nonumber\\
&+&\left.(2n_n+1\pm 1)^{\hf}R^L(k;l_p,n_p,l_n\mp 1,n_n\pm 1)\right\},
\label{25}\er
where $\nu=M\omega/\hbar$ is the oscillator parameter.
We use here $ g_{\sss A} = g_{\sss V}$, a value
almost universaly adopted for the effective axial vector coupling constant,
both in single and double beta nuclear decays.
As in our previous work \cite{Bar96}\footnote{There was a misprint in 
ref.~\cite{Bar96} requiring the following replacement:
\[
{\cal M}_{\sss CM}^{\pm} = 
\sum_{LSJ^{\sss\pi} }m(M_{\sss \pm};L,S,J^{\sss\pi})
\go {\cal M}_{\sss CM}^{\pm} = 
\sum_{LJ^{\sss\pi} }m(M_{\sss \pm};L,S=1,J^{\sss\pi}).
\]}
we made the numerical calculations in the simplifying limit of
 $M_{\sss +}=\infty$, which implies ${\cal M}^+_{\sss CM}=0$, and
$M_{\sss -}=100~\mev$, and we followed
the QRPA procedure described in refs. \cite{Krm96,Krm94}.\footnote{To
assess the reliability of the nuclear model employed here, it is
worthwhile to compare the recently reported result T$_{1/2}^{2\nu} =
(4.3^{+2.4}_{-1.1} [{\rm stat.}] \pm 1.4 [{\rm syst.}]) \times 10^{19}$
y for the $\beta\beta_{2\nu}$ half-life measurement in $^{48}$Ca, with
the predicted value (using the same nuclear structure technique)
T$_{1/2}^{2\nu} = 2.8  \times 10^{19}$ y \cite{Krm94}.  When the
calculation was done only the lower limit of the $2\nu$ decay
half-life,  T$_{1/2}^{2\nu} \geq 3.6 \times 10^{19}$ y, was known
experimentally. On the other hand, the nuclear shell model \cite{Pov95}
restricts the corresponding half-life from above, T$_{1/2}^{2\nu} \leq
10^{20}$ y.}  For the discussion that follows we will only need the
parameters $s$ and $t$, defined as the ratios between the $T=1$, $S=0$
and $T=0$, $S=1$ coupling constants in the particle-particle (PP)
channels and the pairing force  constants, \ie $s=2{\it v}_s^{pp}/[{\it
v}_s^{pair}({\rm p})+{\it v}_s^{pair}({\rm n})]$ and $t=2{\it
v}_t^{pp}/[{\it v}_s^{pair}({\rm p})+{\it v}_s^{pair}({\rm n})]$.  (For
a value of $s \cong 1$ the isospin symmetry is restored within the
QRPA.)

Table \ref{tab1} gives the contributions of the individual matrix
elements in \rf{12} and \rf{13} and their combinations for the charged
majoron $\beta\beta_M$ decay in $^{76}Ge$. Three different results are
shown:  i) the unperturbed or BCS values (second column), ii) the QRPA
calculations when only the particle-hole interaction is considered, \ie
with $s=t=0$ (third column) and iii) the full QRPA calculations with
$s=1$ and $t=1.25$ (fourth column). From the results for the
Gamow-Teller like operators $k^2\mbs_n\cdot\mbs_m/M$ and
$\mbs_n\cdot\kb\mbs_m\cdot\kb/M$ we see that the approximation \rf{11}
is always quite good. This is also true for the individual
contributions, coming both from the natural parity ($\pi=(-)^J$) and
the unnatural parity ($\pi=(-)^{J+1}$) virtual states $J^{\sss\pi}$.
The last two rows show that the approximation \rf{10} is also quite
reasonable. This is clearly because the contributions coming from the
velocity dependent operators, as well as from the Fermi-like operator
$k^2/M$, are relatively small.  It should be noticed that the operators
$-2(\mbs_n\cdot\kb) (\mbs_m\cdot\pb_m)/M$ and $-2i\kb\cdot
\mbs_n\times\pb_m/M$ ($-2\pb_n\cdot\kb/M$ and $-k^2/M$) are forbidden
by the $L$ and $J$ selection rules for the natural (unnatural) parity
intermediate states in eqs. \rf{15}, \rf{16} and \rf{17}.  Moreover,
while all intermediate states contribute coherently for the velocity
independent operators, the same is not true of the velocity dependent
operators. The latter are also more sensitive to the PP correlations
than the former, to the extent that their contributions pass through
zero for $t\leq 1.25$. Table \ref{tab2} shows the comparison of the
exact QRPA results from eq.~\rf{19}, for several $\beta\beta$ decaying
nuclei, with those resulting from the approximation \rf{20}.  We have
used here $s=1$ and $t=t_{\sss sym}$ as discussed in ref.~\cite{Krm94},
\ie $t=1.25,~1.30,~1.50,~1.40$ and $1.85$ for $^{76}Ge$, $^{82}Se$,
$^{100} Mo$, $^{128}Te$ and $^{150}Nd$, respectively.  In all cases the
exact matrix elements are larger in magnitude than the approximate ones
by $\leq 15\%$.

We would also like to point out that our formulation is applicable to
matrix elements that appear in the usual $\beta\beta_{0\nu}$-decay, as
well some supersymmetric contributions. For the former, the
neutrino mass term has been evaluated within this framework in
refs.~\cite{Krm92,Krm94}. In addition, the recoil matrix element
\cite{Doi85}-\cite{Tom91}
\be
M_R^{'(0\nu)}=\left<-i\int \frac{d\kb}{(2\pi)^2}\kb\cdot
(\mbs_n\times{\bf D}_m+ {\bf D}_n\times\mbs_m)
\frac{e^{i\kr_{nm}}}{k(k+\mu)} \right>,
\label{26}\ee
which contributes to neutrinoless $\beta\beta$ decay in models where the
right-handed ($V+A$) exists along with the usual left-handed
($V-A$) one, can be easily calculated from eq. \rf{16} by
making the substitution
\be
g_{\sss A}g_{\sss V} v(k;M_{\pm})\go\frac{1}{\pi}\frac{1}{k(k+\mu)}.
\label{27}\ee
(The finite nucleon size effect and the short-range two-nucleon 
correlations
can also be incorporated in a simple way \cite{Krm94}.\footnote{
So far the matrix element $M_R^{'(0\nu)}$ has been handled as a two-body
operator, which leads to rather complicated analytical expressions
(see eqs. (2.20) to (2.27) in ref.~\cite{Tom86} and eqs.~(3.65) to
(3.68) in ref.~\cite{Tom91}).})  
Furthermore in supersymmetric models
\cite{Hir96b} a matrix element arises with a neutrino potential of
the form
\be
h_R(r)\sim\frac{2}{\pi}\int dk\frac{k^4}{\w(\w+\mu)}
j_0(kr)=\int \frac{d\kb}{2\pi^2}\frac{k^2}{\w(\w+\mu)}
e^{i\kb\cdot\rb},
\ee
whose extra powers of $k^2$ relative to the usual neutrino potential show
that this is similar to the recoil-like terms we evaluated above, 
making this
matrix element also amenable to calculation by our method.

In summary, we have developed a formalism especially suited for
computing nuclear form factors that contain nuclear recoil operators,
regardless of the nuclear model, and we applied it to the matrix
elements for charged majoron emission in the QRPA.  We thereby found
that the approximations used in our previous paper \cite{Bar96} to
simplify the nuclear form factor \rf{6} work quite well, to within a
relative error of 15\%.

\newpage
\begin{table}[t]
\begin{center}
\caption {Contributions of the individual matrix elements
and their combinations for the charged majoron decay in $^{76}Ge$.}
\label{tab1}
\bigskip
\begin{tabular}{|l|rrr|}
\hline
\hline
Operator&BCS&ph-QRPA&full-QRPA\\
\hline
$k^2\mbs_n\cdot\mbs_m/M$&$\!0.200\!$&$\!0.165~~~$&$\!0.112~~~$\\
$\mbs_n\cdot\kb\mbs_m\cdot\kb/M$&$\!0.067\!$&$\!0.055~~~\!$
&$\!0.037~~~\!$\\
$-k^2/M$&$\!0.067\!$&$\!0.049~~~\!$&$\!0.035~~~\!$\\
$-2(\mbs_n\cdot\kb)(\mbs_m\cdot\pb_m)/M$&$\!-0.026\!$&$\!-0.014~~~\!$
&$\!0.007~~~\!$\\
$-2i\kb\cdot \mbs_n\times\pb_m/M$&$\!0.016\!$&$\!0.014~~~\!$&
$\!-0.004~~~\!$\\
$-2\pb_n\cdot\kb/M$&$\!0.006\!$&$\!0.006~~~\!$&$\!0.006~~~\!$\\
$-g_{\sss A}^2\kb\cdot(\mbs_nC_m-C_n\mbs_m)$
&$\!0.041\!$&$\!0.041~~~\!$&$\!0.044~~~\!$\\
$-ig_{\sss V}g_{\sss A}\kb\cdot(\mbs_n\times{\bf D}_m+
{\bf D}_n\times\mbs_m)$
&$\!0.640\!$&$\!0.530~~~\!$&$\!0.348~~~\!$\\
$-g_{\sss V}^2\kb\cdot({\bf D}_n-{\bf D}_m)$
&$\!0.073\!$&$\!0.054~~~\!$&$\!0.041~~~\!$\\
$i\kb\cdot{\bf Y}_{Rnm}$&$\!0.754\!$&$\!0.624~~~\!$&$\!0.433~~~\!$\\
$g_{\sss A}k^2\mbs_n\cdot\mbs_m
( 2g_{\sss V}f_{\sss W}+g_{\sss A})/3M$
&$\!0.694\!$&$\!0.571~~~\!$&$\!0.387~~~\!$\\
\hline \hline \end{tabular} \end{center}
\end{table}
\begin{table}[t]
\begin{center} 
\caption {Results for the exact and approximated matrix elements,
given eqs. \protect \rf{19} and \protect \rf{20}, respectively. Both
the particle-hole and the particle-particle channels have been
considered in the QRPA calculations.}
\label{tab2}
\bigskip
\begin{tabular}{|l|ccccc|}
\hline
\hline
Operator&$^{76}Ge$&$^{82}Se$&$^{100}Mo$&$^{128}Te$&$^{150}Nd$\\
\hline
$i\kb\cdot{\bf Y}_{Rnm}$&$\!0.433\!$&$\!0.437~~~\!$&$\!0.444~~~\!$
&$\!0.412~~~\!$&$\!0.345~~~\!$\\
$g_{\sss A}k^2\mbs_n\cdot\mbs_m
( 2g_{\sss V}f_{\sss W}+g_{\sss A})/3M$
&$\!0.387\!$&$\!0.388~~~\!$&$\!0.380~~~\!$&$\!0.359~~~\!$&$\!0.288~~~\!$\\
\hline \hline \end{tabular} \end{center}
\end{table}
\end{document}